\documentclass[amsmath,amssymb,amsfonts,aps,pre,superscriptaddress,bibnotes,showpacs,showkeys,longbibliography,10pt,twocolumn]{revtex4-2}

\usepackage{graphicx}
\usepackage{dcolumn}
\usepackage{bm}
\usepackage{hyperref}
\usepackage{lipsum}
\usepackage{xcolor}
\usepackage{bbold}
\usepackage{booktabs}
\usepackage{multirow}

\graphicspath{{Figures/}}

\begin{document}

\title{Decoding chirality in circuit topology of a self entangled chain through braiding}

\author{Jonas Berx}
\affiliation{Medical Systems Biophysics and Bioengineering, Leiden Academic Centre for Drug Research, Faculty of Science, Leiden University, 2333CC Leiden, The Netherlands}

\author{Alireza Mashaghi}%
\email{alireza.mashaghi@lacdr.leidenuniv.nl}
\affiliation{Medical Systems Biophysics and Bioengineering, Leiden Academic Centre for Drug Research, Faculty of Science, Leiden University, 2333CC Leiden, The Netherlands}

\date{\today}

\begin{abstract}
Circuit topology employs fundamental units of entanglement, known as soft contacts, for constructing knots from the bottom up, utilizing circuit topology relations, namely parallel, series, cross, and concerted relations. In this article, we further develop this approach to facilitate the analysis of chirality, which is a significant quantity in polymer chemistry. To achieve this, we translate the circuit topology approach to knot engineering into a braid-theoretic framework. This enables us to calculate the Jones polynomial for all possible binary combinations of contacts in cross or concerted relations and to show that, for series and parallel relations, the polynomial factorises. Our results demonstrate that the Jones polynomial provides a powerful tool for analysing the chirality of molecular knots constructed using circuit topology. The framework presented here can be used to design and engineer a wide range of entangled chain with desired chiral properties, with potential applications in fields such as materials science and nanotechnology.

\end{abstract}


\maketitle


\section{\label{sec:intro} Introduction}

Molecular chirality plays a crucial role in biology and soft matter and can generally be classified into chemical, geometrical, and topological chirality \cite{Bonchev2000}. It follows that there are likely couplings between these different types of chirality. Indeed, previous research has shown that chemical chirality or helicity, which refers to the inability of a molecule to switch between enantiomeric configurations by means of intramolecular operations, influences the topological chirality of macromolecular structures, such as polymers. Furthermore, it was recently shown that continuously variable chiral geometries that consist of chemical building blocks with discrete binary chirality can emerge for nano-structured microparticles with bowtie shapes \cite{Kumar2023}. In generic worm-like chain (WLC) models, it was demonstrated that chiral coupling between segments breaks topological mirror symmetry, such that molecular knots formed by closing open-ended chains with a given helicity prefer one macroscopic handedness over the other \cite{Zhao2023}. The ability to tune the chirality in complex molecular configurations is expected to lead to novel innovations in, for example, the design and synthesis of molecular machines \cite{CARPENTER2021,MAYNARD20201914,Horner2016} or the development of chiral photonics \cite{Chen2022}.

Complex knotted chains can be constructed using multiple entangled fundamental building blocks, which we refer to as ``soft contacts”. These chain configurations are created from the bottom up using an approach called circuit topology. By combining different soft contacts in specific ways, it is possible to engineer knots with tailored properties for specific applications. The ability to design and control the chirality of these structures is particularly important, as it can influence their stability, thermodynamics, and response to external stimuli. In these composite systems, there exists an entropic attraction between entangled structures on the same chain \cite{Najafi2016_2,Richard2017}, which can even pass through one-another \cite{Trefz2014}. Experiments with the DNA of a T4 bacteriophage that was stretched in a microfluidic channel using a planar elongational electric field confirmed the theoretical and numerical results related to this entropic attraction \cite{Klotz_2020}. Importantly, the chirality of these structures influences the mutual attraction and the thermodynamics of the macromolecular chain. In particular, a stretched polymer containing two soft contacts possesses a free-energy minimum when both structures are intertwined, and the depth of the minimum is determined by the relative chirality \cite{Najafi2016}. The relative chirality, in turn, determines the stability of the knotted chain configuration.

Circuit topology \cite{golevnev2021} works in 3D and acknowledges the different chiralities that can be present in molecular knots, and encodes chirality in its string notation. To better understand the circuit topology approach, complementary approaches from knot theory can be employed. For example, an Alexander polynomial approach was recently utilised to find parallels between circuit topology operations and knot theory \cite{GOLOVNEV2020}. However, the Alexander polynomial is incapable of distinguishing between enantiomorphic configurations of the same entangled structure. In order to be able to use circuit topology as an effective tool for the characterisation of molecular chains, we need to facilitate the analysis of chirality. In this work, we will do this by calculating the Jones polynomial for binary combinations of soft contacts in all possible circuit topological arrangements. To do so, we will take the approach originally introduced by Jones, i.e., by means of a braid representation into the Temperley-Lieb algebra \cite{Jones1985}. 


\section{\label{sec:scontacts} Soft-contacts: building blocks of molecular knots}
\begin{figure}[htp]
    \centering
    \includegraphics[width=0.9\linewidth]{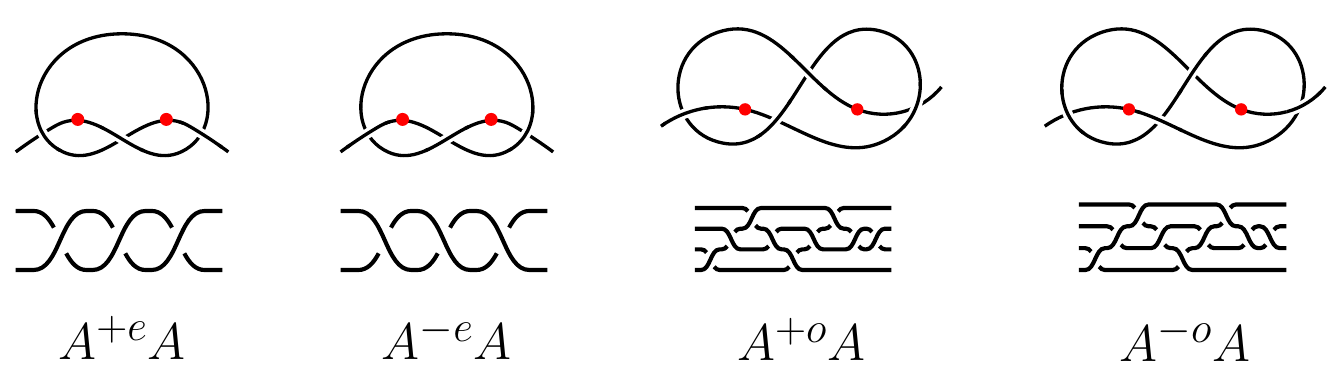}
    \caption{The four fundamental soft contacts needed in the circuit topology framework (top), together with a possible braid representation (bottom). The braid is read from left to right and from bottom to top.}
    \label{fig:soft_contacts}
\end{figure}

Circuit topology, a theory originally developed to investigate the arrangement of contacts in a folded linear chain \cite{Heidari2020}, has recently been generalised to encode chain entanglement \cite{GOLOVNEV2020, GolovnevBook2022}. The smallest structural unit of entanglement in generalised circuit topology (gCT) is the soft contact, or s-contact. This unit is the simplest stable entangled structure that does not change when the chain is deformed. We consider a linear chain that is folded once, and where the end is passed through the resulting loop, creating another loop. There are only four nontrivial configurations that cannot by untied by pulling the ends. One can identify two crossings that form the loops and define a chirality according to the right-hand rule. If both loops have different chirality, the structure will disentangle, and the resulting configuration is not an s-contact. The two remaining crossings define how both loops are locked together; if the chain end is passed through the loops in the same direction as the chirality, we call the resulting s-contact ``even''. When it is passed through in the opposite direction we call the contact ``odd''. We will denote the resulting contacts by a string with superscripts $+e,\,-e,\,+o,\,-o$, depending on the chirality $\{+,-\}$ of the loop and the manner in which the chain end passes through the loop $\{e,o\}$. Only four s-contacts are necessary to construct the complete gCT framework. To make the connection with classical circuit topology (i.e., with hard contacts) more apparent, we will also introduce contact sites, i.e., regions of the chain where the chain gets trapped. Since the soft contacts can be deformed without altering the global structure, the location of these sites is not exact, allowing for a more flexible use of the term. We denote these sites by a red dot in Fig. \ref{fig:soft_contacts}, where we list all possible s-contacts.

The central question is now: how can s-contacts be arranged to construct the topology of a linear chain? Before addressing this question, we first discuss the necessary mathematical tools we will use in this work by means of the $A^{+e}A$ and $A^{-e}A$ s-contacts.


\section{\label{sec:braiding} Braid closures and invariant polynomials}

\begin{figure}[htp]
    \centering
    \includegraphics[width=0.85\linewidth]{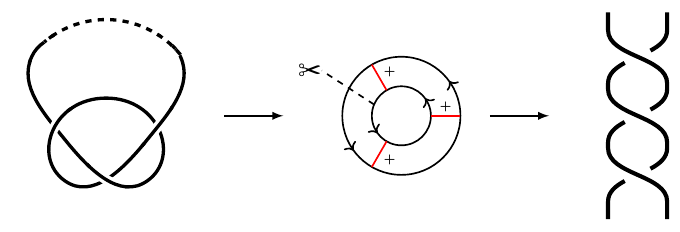}
    \caption{A simple illustration of the Yamada-Vogel algorithm to convert the right-handed trefoil $3_1^+$, corresponding to the closure (denoted by the dashed line) of $A^{+e}A$, into the braid $\beta = \sigma_1^3$.}
    \label{fig:vogel}
\end{figure}

By joining the ends of the $A^{+e}A$ and $A^{-e}A$ contacts without introducing additional crossings, we create the right-handed and left-handed trefoil knots, $K_\pm = \overline{A^{\pm e}A}$, respectively. In Alexander–Briggs notation, these read as $3_1^\pm$, where the superscript denotes the chirality. To proceed, we perform the Yamada-Vogel algorithm \cite{Vogel1990,BIRMAN200519} to turn the knot into a closed braid representation, see Fig. \ref{fig:vogel}.

Since the Yamada-Vogel algorithm is quite involved and depends on the specific sign convention one adheres to, we refer the reader to Ref. \cite{BIRMAN200519} for a detailed discussion and more examples.
Subsequently ``cutting" the closed braid yields a family of algebraic braids that are related to one another by Markov moves, i.e., conjugation and (de-)stabilisation, that can be described by means of a braid word $\beta_K$. The braid word is a string of operators $\sigma_i^{\pm 1}$ that describe whether the $i$-th strand crosses over (positive exponent $+1$) or under (negative exponent $-1$) the $(i+1)$-th strand. For the trefoil knots, the braid word is simply $\beta_{K_\pm} = \sigma_1^{\pm 3}$, where the positive index corresponds to the right-handed trefoil knot $K_+$ and the negative index to the left-handed trefoil knot $K_-$. The writhe $w$ of the braid is then easily found by taking the sum of the exponents in the braid word, i.e., $w_{\pm} = \pm 3$.

\begin{figure}
    \centering
    \includegraphics[width=0.35\linewidth]{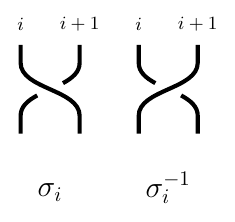}
    \caption{The fundamental braid operators $\sigma_i^{\pm 1}$ indicating over and undercrossings for the $i$-th strand.}
    \label{fig:operators}
\end{figure}

To find polynomials that describe the knot resulting from the braid closure, we will look at two techniques (although others exist that may be simpler): the reduced Burau representation to find the Alexander polynomial $\Delta_K(t)$, and the Kauffmann bracket approach to find the Jones polynomial $J_K(t)$. 

The Burau representation $B$ of a braid with index $n$ is a matrix representation of the operators constituting the braid word. The standard generators $\sigma_i$ of the braid group $B_n$ can explicitly be described in this representation by the matrices
\begin{equation}
    \label{eq:burau}
    \sigma_i\rightarrow \left(\begin{array}{c|cc|c}
        I_{i-1} & 0 & 0 & 0\\
        \hline
        0 & 1-t &t &0\\
        0 & 1 & 0 & 0\\
        \hline
        0 & 0 & 0 & I_{n-i-1}
    \end{array}\right)\,, \qquad i=1,\dots,n-1
\end{equation}
where $0<t\leq 1$ and $I_n$ is the $n\times n$ identity matrix. To find $\sigma_i^{-1}$, it can easily be checked that $\sigma_i$ is invertible by making use of the block diagonal structure. The Burau matrix $B$ for the braid can then be found by matrix multiplication of the generators. Note that since the Burau matrices are row-stochastic, the representation is not irreducible. The generators $\tilde\sigma_i$ in the \emph{reduced} Burau representation are then given by
\begin{equation}
    \label{eq:reduced_burau}
    \begin{split}
    \tilde{\sigma}_1&\rightarrow \left(\begin{array}{cc|c}
        -t & 1 & 0\\
        0 & 1&0\\
        \hline
        0 & 0 & I_{n-3}
    \end{array}\right)\,,\tilde{\sigma}_{n-1}\rightarrow \left(\begin{array}{c|cc}
        I_{n-3} & 0 & 0\\
        \hline
        0 & 1&0\\
        0 & t &-t
    \end{array}\right)\,,\\
    \tilde{\sigma}_i&\rightarrow \left(\begin{array}{c|ccc|c}
        I_{i-2} & 0 & 0 & 0 &0\\
        \hline
        0 & 1 & 0 & 0 & 0 \\
        0 & t & -t & 1 & 0\\
        0 & 0 & 0 & 1 & 0 \\
        \hline
        0 & 0 & 0 & 0 & I_{n-i-2}
    \end{array}\right)\,,\qquad i=1,\dots,n-2\\
    \end{split}
\end{equation}
and for $n=2$, $\tilde\sigma_1 = -t$. The relation between the reduced Burau matrix $\tilde B$ and the Alexander polynomial $\Delta_K(t)$ for the braid closure $K$ is given by
\begin{equation}
    \label{eq:alexander_theorem}
    \Delta_K(t) = \frac{1-t}{1-t^n}\det(I_n - \widetilde B)\,.
\end{equation}

Let us now explicitly perform the calculation for the $A^{+e}A$ and $A^{-e}A$ contacts. Multiplying the matrices corresponding to the reduced operators for $\beta_\pm$, we get the matrix $\tilde B_\pm = -t^{\pm 3}$. The Alexander polynomials are then determined by inserting $\tilde B_\pm$ into equation \eqref{eq:alexander_theorem}, i.e., 
\begin{equation}
    \label{eq:alexander_trefoil}
    \Delta_{K_\pm}(t) = \begin{cases}
        1-t+t^2 & \text{for $K_+$}\\
        t^{-3} -t^{-2} +t^{-1} & \text{for $K_-$}\,.
    \end{cases}
\end{equation}
Although both Laurent polynomials may seem to be different, the Alexander polynomials are only unique up to multiplication by the Laurent monomial $\pm t^{n}$. If we make the choice to normalise $\Delta_{K_\pm}(t)$ in such a manner that the constant term is positive, we can easily see that both Laurent polynomials are, in fact, equal. The consequence of this result is that the Alexander polynomial is insufficient to distinguish between the different chiralities for soft contacts. 

We now shift our attention to the Jones polynomial. Although this polynomial can be calculated easily by means of the Kauffmann bracket algorithm applied directly to the planar projection of the trefoil knots, we want to keep the braid representation as our starting point. To proceed, we define the homomorphism on $n$ strands $\rho_n: B_n \rightarrow \mathrm{TL}_n$ between the braid group $B_n$ and the Temperley-Lieb algebra $\mathrm{TL}_n$ over the ring $\mathbb{Z}[A,A^{-1}]$ as follows:
\begin{equation}
    \label{eq:TL_homomorphism}
    \begin{split}
        \rho_n(\sigma_i) &= A \mathbb{1}_n + A^{-1} U_i\,,\\
        \rho_n(\sigma_i^{-1}) &= A^{-1} \mathbb{1}_n + A U_i\,,
    \end{split}
\end{equation}
where the $U_i$ are the generators of the TL algebra. By mapping the braid word $\beta_K$ to a multiplication of factors from \eqref{eq:TL_homomorphism} and using the Jones relations for multiplications of $U_i$, we find a polynomial in $U_t$, where $t$ indexes all fundamental generators in the Temperly-Lieb algebra. The number of such generators for a braid of $n$ strands is equal to the Catalan number $C_n = {2n \choose n}/(n+1)$. The coefficient of $U_t$ is defined as $\langle \beta_K|t\rangle$, such that the expression for the braid becomes 
\begin{equation}
    \label{eq:braid_TL}
    \rho_n(\beta_K) = \sum_t \langle \beta_K|t\rangle U_t\,.
\end{equation}
We define the bracket for an operator $U_t$ as $\delta^{||U||}$, where $\delta = -A^2 - A^{-2}$ and $||U||$ is the number of components in the unlink minus one, obtained by taking the closure of the Temperley-Lieb diagrams corresponding to $U_t$. A detailed discussion on the algorithm can be found in \cite{Kauffman2013}. 

The normalised bracket polynomial, denoted by $\langle .\rangle$ for the braid representation $\beta_\pm$ of the even $A^{\pm e}A$ contacts, i.e., $\langle\beta_\pm\rangle$ is then
\begin{equation}
    \label{eq:bracket_trefoil}
    \langle\beta_\pm\rangle = A^{\mp 4} + A^{\mp 12} - A^{\mp 16}\,.
\end{equation}
By inserting $A = t^{-1/4}$, we finally get the Jones polynomials, 
\begin{equation}
    \label{eq:jones_trefoil}
    J_{K_\pm} (t) = \begin{cases}
        t+t^3 - t^4 & \text{for $K_+$}\\
        t^{-1} + t^{-3} - t^{-4} & \text{for $K_-$}\,,
    \end{cases}
\end{equation}
where it can now easily be seen that $J_{K_+}(t) = J_{K_-}(1/t)$. Since the Jones polynomial is already normalised, there is no freedom in choosing a monomial prefactor, making the polynomials unique. Hence, the Jones polynomial can differentiate between different chiralities. Note that the Jones polynomial can only be used as proof of the chirality of a knot, not as proof of amphichirality.
\begin{table}[htp]
    \centering
    \caption{The knots resulting from the closure of the fundamental s-contacts. Green and red colours indicate that knots have opposite chirality.}
    \begin{tabular}{ccc}
    \midrule
    $A^{+e}A$ & $A^{-e}A$ & $A^{\pm o}A$\\
        \toprule
        \includegraphics[width=0.2\linewidth]{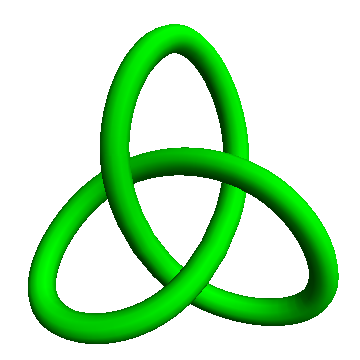}  & \includegraphics[width=0.2\linewidth]{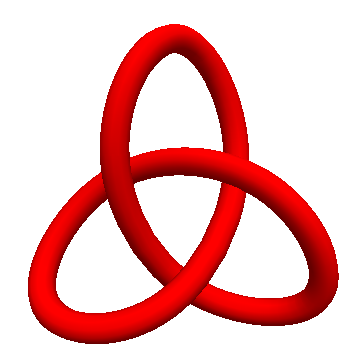} & \includegraphics[width=0.2\linewidth]{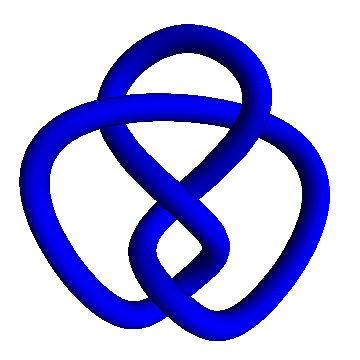}\\
        $3_1^+$ & $3_1^-$ & $4_1$\\
        \bottomrule
    \end{tabular}
    \label{tbl:s-contacts}
\end{table}

What about the odd s-contacts? Repeating the previous calculations for the knots $K_\pm = \overline{A^{\pm o}A}$  resulting from the closures of the corresponding s-contacts yields the Alexander polynomial
\begin{equation}
    \label{eq:alexander_eight}
    \Delta_{K_\pm}(t) = 1-3t+t^2\,,
\end{equation}
for both s-contact chiralities. This was the expected result, as positive and negative chiralities are simply reflections of the contact in the plane. The Jones polynomials are
\begin{equation}
    \label{eq:jones_eight}
    J_{K_\pm} (t) = t^2 -t +1 - t^{-1} + t^{-2}\,.
\end{equation}
They are also the same. This of course makes sense, since the closure of the $A^{\pm o}A$ contacts is the figure-eight knot ($4_1$), which is known to be amphichiral, i.e., capable of being continuously deformed into its own reflection in the plane.

We list the knots obtained by closing the different fundamental s-contacts in Table \ref{tbl:s-contacts}. The chirality associated with the closure of a specific s-contact is given by its colour: green indicates that the sign of the highest-valued power (in absolute value) in the corresponding Jones polynomial is \emph{positive}, while red indicates that it's \emph{negative}. Blue indicates that the resulting closure is \emph{amphichiral}. Red and green colours hence pertain to similar knots with opposite chiralities.  



\section{Generalised circuit topology \label{sec:gCT}}
\begin{figure}[htp]
    \centering
    \includegraphics[width=0.8\linewidth]{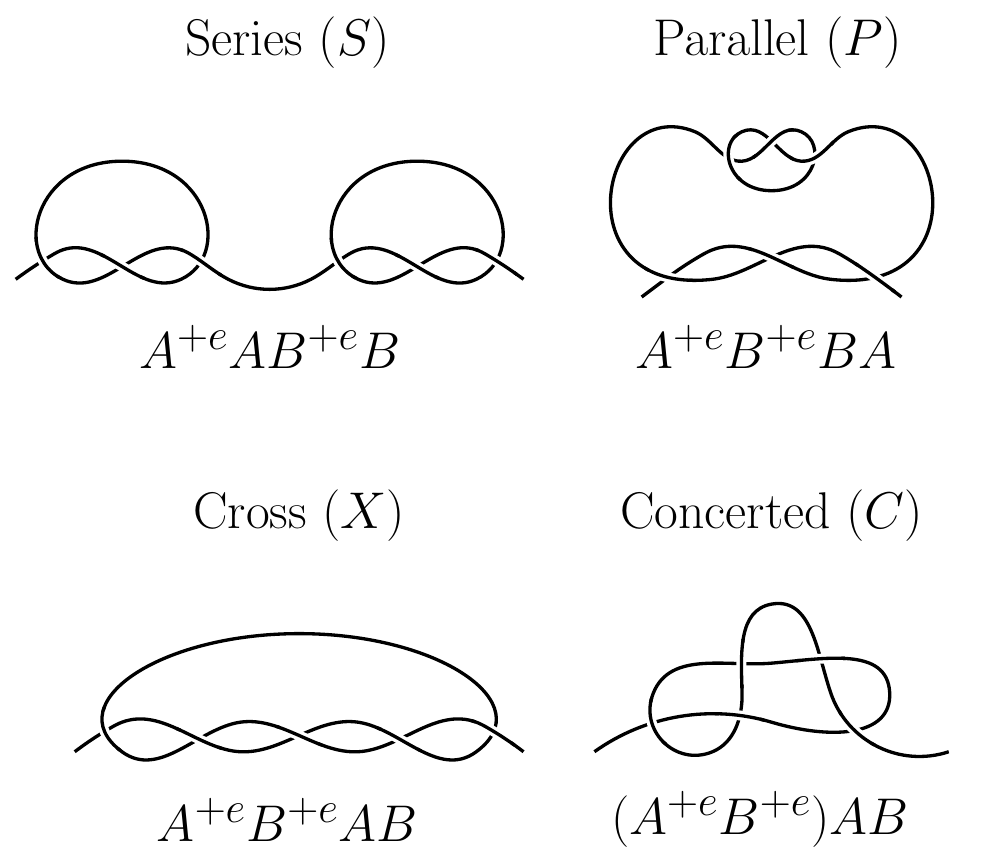}
    \caption{Series $(S)$, parallel $(P)$, cross $(X)$ and concerted $(C)$ configurations of two $A^{+e}A$ soft contacts, together with the corresponding string notation.}
    \label{fig:SPX}
\end{figure}

We now return to circuit topology's central question: how can s-contacts be combined to build all known molecular knots? The CT framework focuses on pairwise relations between contacts $A$ and $B$. Only three relations can be defined for soft contacts: the series ($S$), parallel ($P$) and cross ($X$) configurations, with corresponding string notation $AABB$, $ABBA$ and $ABAB$, respectively. A fourth category is possible if one allows for one shared contact site, the so-called concerted $C$ contacts, with string notation $(AB)AB$. These relations are visualised in Fig. \ref{fig:SPX}.

Let us start by considering the $S$ and $P$ configurations. Suppose we concatenate two $A^{+e}A$ contacts (the series $S$ configuration in the language of CT), resulting in $A^{+e}A B^{+e}B$ and subsequently close the ends. The resulting knot $K'_+ = \overline{A^{+e}A B^{+e}B}$ is in fact the connected sum of the two individual knots, i.e., $K'_+ = K_+\,\#\, K_+$, where $K_+ = \overline{A^{+e}A}$, and hence the Alexander polynomial $\Delta_{K'_+}(t)$ is the product of the polynomials of the constituent knots, i.e., 
\begin{equation}
    \label{eq:alexander_product}
    \Delta_{K'_+}(t) = \Delta_{K_+}^2(t) = 1-2t + 3t^2-2t^3 +t^4\,,
\end{equation}
which is the so-called granny knot. If, however, one performs a similar calculation for a series combination of two $A^{-e}A$ contacts, i.e., for $A^{-e}A B^{-e}B$, or a series combination of $A^{+e}A$ and $A^{-e}A$, i.e., for $A^{+e}A B^{-e}B$, the same Alexander polynomial as in equation \eqref{eq:alexander_product} arises. For the $A^{-e}A B^{-e}B$ arrangement this does not present a problem, since it is again a reflection of the $A^{+e}A B^{+e}B$ series contacts in the plane. The series combination $A^{+e}A B^{-e}B$ (i.e., the square knot upon closure), however, cannot be related to the other configurations in any manner, and so the Alexander polynomial is insufficient to resolve this degeneracy. 

Conversely, since the Jones polynomial can also be factored for connected sums, for the knot $K'_+ = \overline{A^{+e}A B^{+e}B}$ it becomes 
\begin{equation}
    \label{eq:jones_series_trefoil_positive}
    J_{K'_+}(t) = J^2_{K_+}(t) = t^8-2 t^7+t^6-2 t^5+2 t^4+t^2\,,
\end{equation}
while for the knot $K'_- = \overline{A^{-e}A B^{-e}B}$ it becomes 
\begin{equation}
    \label{eq:jones_series_trefoil_negative}
    J_{K'_-}(t) = J^2_{K_-}(t) = t^{-8}-2 t^{-7}+t^{-6}-2 t^{-5}+2 t^{-4}+t^{-2}\,.
\end{equation} 
The mixed series combination $K'_0 = \overline{A^{+e}A B^{-e}B}$ then results in a Jones polynomial
\begin{equation}
    \label{eq:jones_series_trefoil_combo}
    J_{K'_0}(t) = -t^3 +t^2 -t +3 -t^{-1} +t^{-2} -t^{-3}\,.
\end{equation} 
We immediately see from equations \eqref{eq:jones_series_trefoil_positive} - \eqref{eq:jones_series_trefoil_combo} that the Jones polynomials are all different, and that $J_{K'_+}(t) = J_{K_-}(t^{-1})$.

Note that we did not consider the $P$ configuration of contacts. This is because the $S$ and $P$ configurations are identical. One can ``shrink'' one of the two soft contacts in a series arrangement and slide it through the other contact, until it is located in the loop. Expanding the contact again results in an overall parallel arrangement of the circuit \cite{Klotz_2020}. Hence, both configurations are described by a connected sum between closures of the constituent s-contacts. A good example of this equivalence is given in \cite{golevnev2021}. In fact, it can be shown that the chain entropy is maximal when the two soft contacts are intertwined, i.e., in the $P$ configuration, where the distance between the centres of mass is minimal.

With the polynomial invariants from the closures of the even and odd s-contacts, one can construct any arrangement of series or parallel configurations. The polynomials for the cross and concerted configurations, however, are not trivially found by taking connected sums. In the next section, we will devote our attention to these topological arrangements.


\section{\label{sec:cross_concert}The cross and concerted configurations}
We now consider the nontrivial combinations of s-contacts, i.e., the cross ($X$) and concerted ($C$) configurations. Although they may seem similar, there are crucial differences between them; $C$ configurations are only nontrivial for some combinations of s-contacts while nontrivial $X$ configurations are always possible to construct.
\subsection{Cross contacts}
For the cross configuration, we will consider all 16 combinations of $A^{\pm e/o}B^{\pm e/o}AB$, which are not necessarily distinct, see Table \ref{tbl:cross_knots}. 
\begin{table}[htp]
    \centering
    \caption{The knots resulting from the closure of cross-contacts with different chiralities. Green and red colours indicate that knots have opposite chirality, while blue knots are amphichiral.}
    \begin{tabular}{ccccc}
        \toprule
        {} & $B^{+e}$ & $B^{-e}$ & $B^{+o}$ & $B^{-o}$ \\
        \midrule
        $A^{+e}$ & \includegraphics[width=0.2\linewidth]{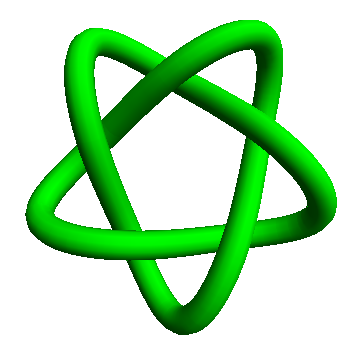}  & \includegraphics[width=0.2\linewidth]{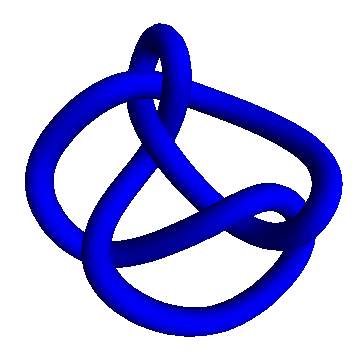} & \includegraphics[width=0.2\linewidth]{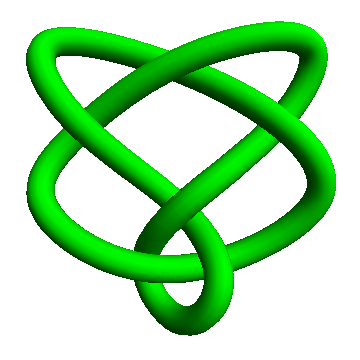} & \includegraphics[width=0.2\linewidth]{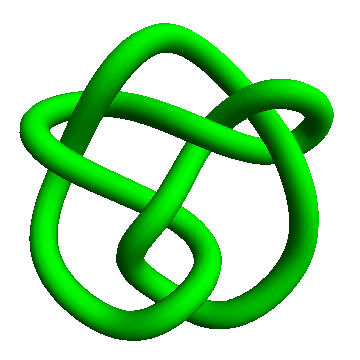}\\
        {}& $5_1^+ $ & $6_3 $ & $6_2^+ $ & $7_6^+ $\\
        $A^{-e}$ & \includegraphics[width=0.2\linewidth]{6_3.png} & \includegraphics[width=0.2\linewidth]{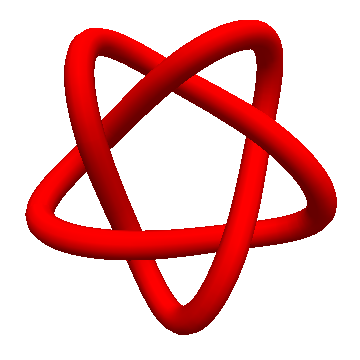} & \includegraphics[width=0.2\linewidth]{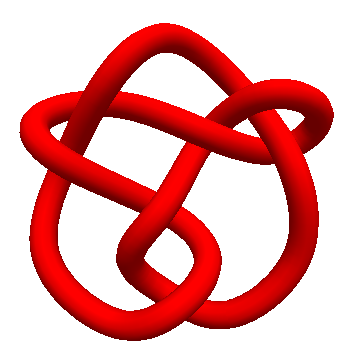} & \includegraphics[width=0.2\linewidth]{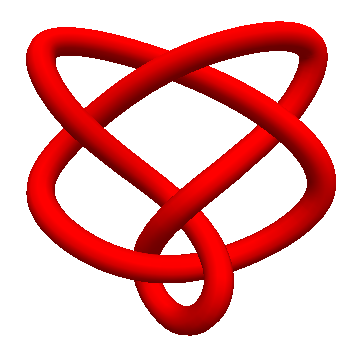} \\
        {}& $6_3 $ & $5_1^- $ & $7_6^-$ & $6_2^-$\\
        $A^{+o}$ & \includegraphics[width=0.2\linewidth]{6_2bar.png} & \includegraphics[width=0.2\linewidth]{7_6.png} & \includegraphics[width=0.2\linewidth]{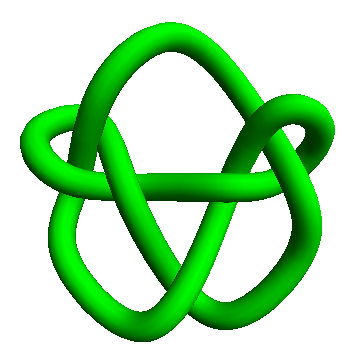} & \includegraphics[width=0.2\linewidth]{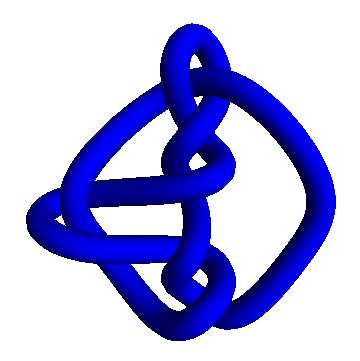}\\
        {}& $6_2^- $ & $7_6^+ $ & $7_7^+ $ & $8_{12} $\\
        $A^{-o}$ & \includegraphics[width=0.2\linewidth]{7_6bar.png} & \includegraphics[width=0.2\linewidth]{6_2.png} & \includegraphics[width=0.2\linewidth]{8_12.png} & \includegraphics[width=0.2\linewidth]{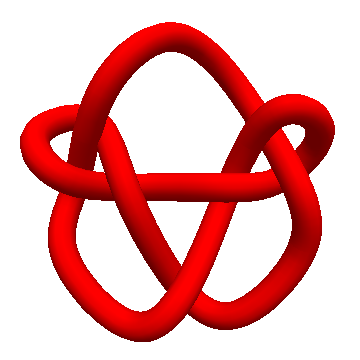}\\
        {}& $7_6^- $ & $6_2^+$ & $8_{12} $ & $7_7^- $\\
        \bottomrule
    \end{tabular}
    \label{tbl:cross_knots}
\end{table}
Let us consider a simple example for illustration purposes: the $A^{+e}B^{+o}AB$ cross configuration. By closing this configuration and performing, e.g., Vogel's algorithm, we get the knot $K$ with possible braid representation $\beta_K = \sigma_3\sigma_2^4 \sigma_1\sigma_2^{-1}\sigma_3^{-1}\sigma_2^{-1}\sigma_1^{-1}\sigma_2^{-1}$ on four strands. The resulting Alexander and Jones polynomials are 
\begin{equation}
    \label{eq:6_2_polynomials}
    \begin{split}
        \Delta_K(t) &= 1-3t+3t^2-3t^3+t^4\,,\\
        J_K(t) &= t^{-1} -1 +2t-2t^2+2t^3-2t^4+t^5\,,
    \end{split}
\end{equation}
which leads to the conclusion that the closure of $A^{+e}B^{+o}AB$ results in the $6_2^+$ prime knot (or Miller Institute knot). Its chiral opposite $6_2^-$ can be found by either flipping the chirality of the entire contact, i.e., $A^{-e}B^{-o}AB$, or by switching the order, i.e., $A^{+o}B^{+e}AB$. Applying both operations on the contact yields $A^{-o}B^{-e}AB$ and results in the exact same prime knot $6_2^+$.

A similar line of reasoning can be applied to every configuration; when the two s-contacts are flipped, the chirality of the cross contact changes; this amounts to reversing the orientation in which we read the contact. When the chirality of both components are flipped simultaneously, the total chirality of the cross contact also flips. Therefore, we can construct Table \ref{tbl:cross_knots} by only considering a small subset of contacts and derive the others by symmetry arguments. The table is symmetric up to a change of the total chirality, so this leaves 10 configurations to consider. However, since flipping the chirality of both contacts flips the total chirality, the contacts $A^{+e}B^{+e}AB$ and $A^{-e}B^{-e}AB$ also yield the same knot, but with different chirality. Similar for $A^{+o}B^{+o}AB$, $A^{+e}B^{+o}$, and $A^{+e}B^{-o}$. This leaves us with only six independent configurations that need to be checked, reducing the number of computations required.

\subsection{Concerted contacts}
In analogy with circuit topology for hard contacts, we define a concerted contact to be the contact formed by assuming that a contact site is shared, i.e., the two s-contacts share a single loop. We denote a concerted contact in a similar manner as a cross contact, but with brackets around the shared contact pair, e.g., $(A^{+e}B^{+e})AB$, which is different from the cross contact $A^{+e}B^{+e}AB$. Since both contacts share a single loop, their chirality must be identical, otherwise the structure would disentangle. If one were to try to create such a knot, it would lead to so-called \emph{slip-knots}, which are isotopic to the unknot. This then leaves us with eight nontrivial options. Moreover, it is also easily seen that concerted contacts consisting of both an even and an odd contact result in the unknot. The only remaining possibilities are then the contacts $(A^{+e}B^{+e})AB$, $(A^{+o}B^{+o})AB$ and their chiral opposites which we list in Table \ref{tbl:concerted_knots}. To simplify notation, we introduce the following abbreviation $(A^{+e}B^{+e})AB \rightarrow A^{+2e}A$, where the exponent indicates the number of concerted contacts.
\begin{table}[htp]
    \centering
    \caption{The knots resulting from the closure of nontrivial concerted contacts with different chiralities. Green and red colours indicate that knots have opposite chirality.}
    \begin{tabular}{cccc}
    \toprule
    $A^{+2e}A$ & $A^{-2e}A$ & $A^{+2o}A$ & $A^{-2o}A$\\
        \midrule
        \includegraphics[width=0.2\linewidth]{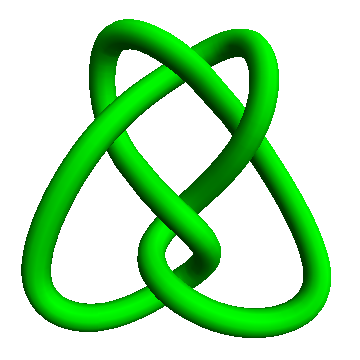}  & \includegraphics[width=0.2\linewidth]{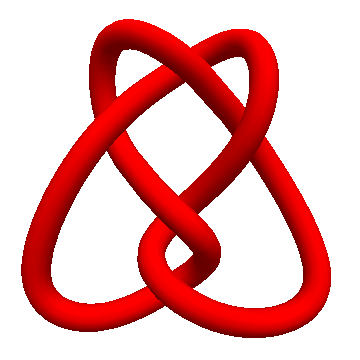} & \includegraphics[width=0.2\linewidth]{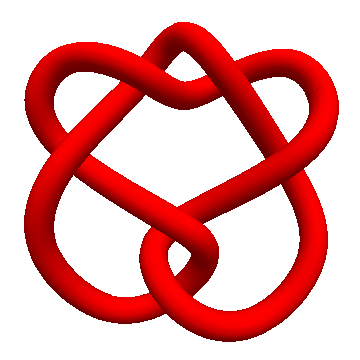} & \includegraphics[width=0.2\linewidth]{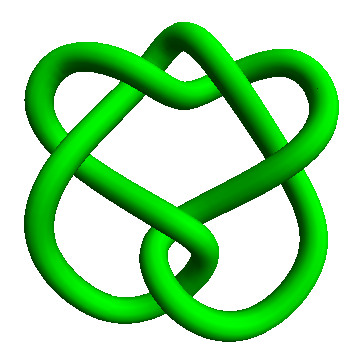}\\
        $5_2^+$ & $5_2^-$ & $6_1^-$ & $6_1^+$\\
        \bottomrule
    \end{tabular}
    \label{tbl:concerted_knots}
\end{table}
The closure $K$ of the $A^{+2e}A$ concerted contact can be reshaped into a braid with word $\beta = \sigma_2^2\sigma_3\sigma_2\sigma_1\sigma_2\sigma_3^{-1}\sigma_2\sigma_1^{-1}$. The associated Alexander and Jones polynomials are 
\begin{align}
    \Delta_K(t) &= 2-3t+2t^2\label{eq:concerted_+2e_alexander}\\
    J_K(t) &= t-t^2+2t^3-t^4+t^5-t^6\,,\label{eq:concerted_+2e_jones}
\end{align}
while for the chiral opposite $K' = \overline{A^{-2e}}$, the Jones polynomial is $J_{K'}(t) = J_K(1/t)$.


\section{Building circuits\label{sec:extend}}
Of course, in more complex molecules, more than two s-contacts need to be considered. The modular arrangement of all of these s-contacts constitutes a \emph{circuit}, which can be described by judiciously combining them in any of the aforementioned configurations $(S,\,P,\,X,\,C)$.

Again, for the $S$ and $P$ configurations, this does not present a problem, since the Jones polynomial factorises into simpler polynomials, but for consecutive $X$ or $C$ configurations we require a general rule. Let us first consider concerted topologies, since they are somewhat simpler. The closure of the concerted s-contacts $(A^{+e}B^{+e}C^{+e})ABC$, or $A^{+3e}A$ for simplicity, yields a configuration where one end of the chain is looped three times around the large loop. This results in a twist knot with five half-twists and positive chirality. We quickly notice a pattern here; if the concerted contact consists of $k\in\mathbb{N}$ positive even s-contacts, i.e., $A^{+k e}A$, the resulting closure will be a twist knot with $n=2k-1$ half-twists, and the Jones polynomial will consequently be 
\begin{equation}
    \label{eq:twist_knot_jones_even}
    J_{A^{+ke}A}(t) = \frac{t+t^3 + t^{2k} - t^{3+2k}}{t+1}\,.
\end{equation}
Flipping the chirality of all the s-contacts then gives us that $J_{A^{-ke}A}(t) = J_{A^{+ke}A}(t^{-1})$. The odd contacts $A^{+ko}A$ give a twist knot with an even number of half-twists $n=2k$, resulting in the Jones polynomial
\begin{equation}
    \label{eq:twist_knot_jones_odd}
    J_{A^{+ko}A}(t) = \frac{t+t^3 + t^{-2k} - t^{3-2k}}{t+1}\,.
\end{equation}

What about more exotic combinations of $C$ contacts? It can easily be shown that in a string of $C$ contacts, one can look at binary combinations of pairs of contacts. Since pairs that alternate between even and odd always untie, a string of concerted contacts can be reduced to a single twist knot with Jones polynomial given by equation \eqref{eq:twist_knot_jones_even} or \eqref{eq:twist_knot_jones_odd}. For example, the configuration $(A^{+e} B^{+3o} C^{+e})ABC$ reduces to the single s-contact $B^{+o}B$ by first eliminating the $(A^{+e}B^{+o})AB$ contact, and then by eliminating $(B^{+o}C^{+e})BC$.

For cross contacts, the computations are significantly more difficult. We can recognise that combining three identical s-contacts in an $X$ configuration, e.g., $A^{+e}B^{+e}C^{+e}ABC$ yields a torus knot $T_{p,q}$, where $p$ and $q$ are coprime integers, with an odd number of crossings. The Jones polynomial of a torus knot is
\begin{equation}
    \label{eq:jones_torus}
    J(t) = t^{(p-1)(q-1)/2}\left(\frac{1-t^{p+1}-t^{q+1}+t^{p+q}}{1-t^2}\right)\,.
\end{equation}
The $p$ and $q$ indices indicate the number of crossings and the number of strands in the closed braid representation, respectively. It can be shown graphically that $q=2$ for all torus knots obtained in the manner described above. We will henceforth also denote the $k-$fold cross contact as $T_{\pm (2k+1),2}$, where the sign corresponds to the global chirality of the contact. As a check, we can confirm that the $2-$fold $A^{+e}B^{+e}AB$ contact can be written as $T_{5,2}$, which is the $5_1^+$ knot, and that, e.g., $A^{-e}B^{-e}C^{-e}ABC \rightarrow T_{-7,2}$, which is the $7_1^-$ knot.

Twist and torus knots are of the utmost importance in the study of DNA knots. Of these two knot families, twist knots are more commonly found in nature since their unknotting number $u$ is always equal to one, while torus knots are characteristically over-represented in experiments with viral DNA due to specific properties of DNA packing \cite{Arsuaga2005,Micheletti2022}. The unknotting number for the $k-$fold cross contact equals $u = (p-1) (q-1)/2 = k$, i.e., the minimum number of times the strand must be passed through itself to untie it exactly equals the number of identical s-contacts that constitute the circuit. Hence, we can expect concerted contacts to be ubiquitous in applications involving more complicated circuits.

Since there exists only one torus knot for each crossing number, we can immediately deduce that knots formed from other s-contact combinations must be described by other rules. To the best of our knowledge, we do not know whether such a framework exists, and hence we leave it for future research.

\begin{figure}[htp]
    \centering
    \includegraphics[width=0.6\linewidth]{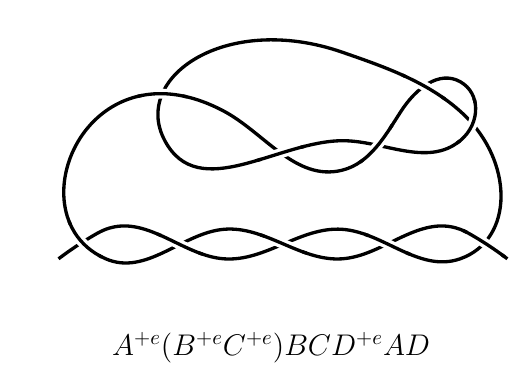}
    \caption{The circuit formed by combining a $C$, $P$ and $X$ configuration, resulting in the string $A^{+e}(B^{+e}C^{+e})BCD^{+e}AD$. It can easily be seen that the top part of the circuit (i.e., the concerted $B^{+2e} = (B^{+e}C^{+e})BC$ relation) is in a $P$ relation with the $A^{+e}D^{+e}AD$ subcircuit, which in turn is a cross relation between $A^{+e}A$ and $D^{+e}D$.}
    \label{fig:CPX_circuit}
\end{figure}

For circuits involving more complex combinations of s-contacts, the modularity of those contacts can be used to derive Jones polynomials. An example is given in Fig. \ref{fig:CPX_circuit}, where the circuit $A^{+e}(B^{+e}C^{+e})BCD^{+e}AD$ is shown; for simplicity we have chosen only the $+e$ chirality for all contacts. It can be easily seen from the string notation that contacts $B$ and $C$ are in a concerted relation and can hence be simplified to $(B^{+e}C^{+e})BC = B^{+2e}$, where the equality sign denotes ambient isotopic equivalence of the corresponding knot closure. The concerted contact itself is now in a parallel relation with the $A^{+e}A$ contact, which is in turn in a cross relation with the $D^{+e}D$ contact. Since the $B^{+2e}$ configuration is in parallel with the rest of the circuit, it can be moved out of the string notation, i.e., $A^{+e}(B^{+e}C^{+e})BCD^{+e}AD = A^{+e}D^{+e}AD B^{+2e}$. The resulting Jones polynomial of the entire circuit is then the product of the Jones polynomials of the cross contact $A^{+e}D^{+e}AD$ and the concerted contact $B^{+2e}$, i.e., from Table \ref{tbl:jones_list} in the appendix \ref{app:A1},
\begin{equation}
    \label{eq:jones_circuit}
    \begin{split}
    J(t) &= J_{A^{+e}D^{+e}AD}(t) \times J_{B^{+2e}}(t)\\
    &= \left(-t^{7} + t^{6} -t^{5} +t^{4} +t^{2}\right)\\
    &\times\left(-t^{6}+t^{5}-t^{4}+2t^{3}-t^{2}+t\right)\,.
    \end{split}
\end{equation}
This approach is quite general; for circuits that involve \emph{distinct} $S,\,P$ configurations, among others, one can separate the different subcircuits. The calculation of the Jones polynomial hence reduces to the individual computation of the subcircuits' polynomials. However, when the $S$ or $P$ contacts are in some sense ``blocked'' by others, e.g., in the circuit $ACABCB$, they cannot be separated. Contacts $A$ and $B$ are connected in series, but they cannot be connected in parallel due to their inability to be pulled along the string. The dragging motion is hindered by contact $C$, which intersects with contacts $A$ and $B$. A circuit can then be defined as a distinct section of a string that comprises exclusively of pairs of letters \cite{GOLOVNEV2020}. In essence, a circuit can be separated (put in series) or distinguished from other contacts and circuits. No general theory to compute the Jones polynomial of such inseparable circuits has been found (and is unlikely to exist); we hence leave this for future research.


\section{Conclusions \label{sec:conclusions}}

Circuit topology, a theory that describes intra-chain contacts in folded open chains, was recently generalised to account for chain entanglement \cite{GOLOVNEV2020}. This unique generalisation, specific to circuit topology and not derived from other theories, allows for the construction of various types of knots, including those commonly found in biomolecules \cite{golevnev2021, Flapan2023-hx, JACKSON2017,Flapan2019,CARPENTER2021}. The approach reveals patterns not apparent from traditional knot theoretic approaches \cite{golevnev2021}. For example, protein knots form a well-defined, distinct group, which naturally appears if expressed in terms of circuit topology units and operations. In this paper, we have studied the advantages of using a circuit topology framework for engineering of molecular knots, and paid special attention to the chirality of the structures by means of a braid-theoretic framework. Our approach demonstrates the importance of considering chirality in the construction of topological circuits, and show that the Jones polynomial is an effective invariant for characterising their properties. 

The presented approach allows us to set the stage for future studies on (braided) multichain systems that can exhibit more complex intra- and interchain entanglement, and hence can contain links, necessitating the use of invariant polynomials beyond the Alexander polynomial. By expanding the scope of our framework, we can gain deeper insights into the structure and function of biomolecules and other complex materials, paving the way for the development of new technologies and applications in fields such as biophysics, materials science, and nanotechnology.

\appendix
\section{\label{app:A1}}
We present a list of the Jones polynomials for all s-contacts, as well as their combinations using cross and concerted relations, together with the knot and associated unknotting number they represent when closed. 

\begin{table*}[pt]
    \centering
    \caption{The Jones polynomials $J(t)$ and knots obtained from the closure of several soft contact configurations. The unknotting number $u$ is also given in the final column.}
    \begin{tabular}{cccc}
        \toprule
         Configuration & Jones polynomial & Knot & Unknotting nr. \\
        \midrule
        $A^{\pm e}A$ & $- t^{\pm 4} + t^{\pm 3} +t^{\pm 1}$ & $3_1^\pm$ & $1$\\
        $A^{\pm o}A$ & $t^2 -t +1 - t^{-1} + t^{-2}$ & $4_1$ & $1$\\
        \midrule
        $A^{\pm 2e}A$ & $-t^{\pm 6}+t^{\pm  5}-t^{\pm 4}+2t^{\pm 3}-t^{\pm 2}+t^{\pm 1}$ & $5_2^\pm$ & $1$\\
        $A^{\pm 2o}A$ & $t^{\mp 4} - t^{\mp 3} + t^{\mp 2} -2t^{\mp 1}-2-t^{\pm 1}+t^{\pm 2}$ & $6_1^\mp$ & $1$\\
        \vdots & \vdots & \vdots\\
        \midrule
        $A^{\pm e}B^{\pm e}AB$ & $-t^{\pm 7} + t^{\pm 6} -t^{\pm 5} +t^{\pm 4} +t^{\pm 2}$ & $5_1^\pm$ & $2$\\
        $A^{\pm e}B^{\mp e}AB$ & $-t^3 +2t^2-2t+3-2t^{-1}+2t^{-2}-t^{-3}$ & $6_3$ & $1$\\
            $A^{\pm e/o}B^{\pm o/e}AB$
        & $t^{\pm 5} - 2t^{\pm 4}+2t^{\pm 3}-2t^{\pm 2}+2t^{\pm 1}-1+t^{\mp 1}$ & $6_2^\pm$ & $1$\\
            $A^{\pm e/o}B^{\mp o/e}AB$
        & $-t^{\pm 6} +2t^{\pm 5} - 3t^{\pm 4}+4t^{\pm 3}-3t^{\pm 2}+3t^{\pm 1}-2+t^{\mp 1}$ & $7_6^\pm$ & $1$\\
        $A^{\pm o}B^{\pm o}AB$ & $t^{\pm 4}-2t^{\pm 3}+3t^{\pm 2}-4t^{\pm 1}+4-3t^{\mp 1}+3t^{\mp 2}-t^{\mp 3}$ & $7_7^\pm$ & $1$\\
        $A^{\pm o}B^{\mp o}AB$ & $t^4 - 2t^3+4t^2-5t+5-5t^{-1}+4t^{-2}-2t^{-3}+t^{-4}$ & $8_{12}$ & $2$\\
        \midrule
        $A^{\pm e}B^{\pm e}C^{\pm e}ABC$ & $-t^{\pm 10} + t^{\pm 9} - t^{\pm 8} + t^{\pm 7} - t^{\pm 6} + t^{\pm 5} + t^{\pm 3}$ & $7_1^\pm$ & $3$
    \end{tabular}
    \label{tbl:jones_list}
\end{table*}

\bibliographystyle{apsrev4-2}
\bibliography{apssamp}

\end{document}